\documentstyle{elsart}

\begin{document}

\mbox{} \hfill UCRHEP-T249\\
\mbox{} \hfill EFUAZ FT-99-68\\

\begin{frontmatter}

\title{\bf $\mathbf L/E$-Flatness of the Electron-Like Event Ratio\\
                    in Super-Kamiokande and a Degeneracy\\
                    in Neutrino Masses}

\author{I. Stancu$^a$ and D. V. Ahluwalia$^b$}

\address{ $^a$ Department of Physics, University of California\\
Riverside, CA 92521, USA \\
E-mail: ion.stancu@ucr.edu\\ 
\vskip 0.25cm
$^b$ Escuela de Fisica, Univ. Aut. de Zacatecas\\
Apartado Postal C-580, Zacatecas 98068, Mexico\\
E-mail: ahluwalia@phases.reduaz.mx, av@p25hp.lanl.gov}


\vfill
\begin{abstract}
We show that the $L/E$-flatness of the electron-like event ratio in the 
Super-Kamiokande atmospheric neutrino data implies the equality of the
expectation values for the muon and tau neutrino masses.
We establish this result by obtaining a set of three constraints on the
neutrino-oscillation mixing matrix as contained in the indicated flatness.
The resulting $3\times 3$ neutrino-oscillation matrix depends only on one
angle.
A remarkable result that follows directly from this matrix is the consistency
between the mixing angles observed by LSND and Super-Kamiokande.
\end{abstract}

\centerline{{\bf Journal Ref.:} Phys. Lett. B 460, 431-436 (1999)}

\end{frontmatter}
\newpage
%
%
\def\beq{\begin{eqnarray}}
\def\eeq{\end{eqnarray}}
%
%
\section{Introduction}

One of the most remarkable features of the latest Super-Kamiokande
data on the atmospheric neutrino anomaly is the $L/E$-flatness
of the electron-like event ratio~\cite{SuperK}.
Given the definitions,
\beq
{\cal R}_e &\equiv& \frac{\mbox{Experimentally observed $e$-like events}}
                    {\mbox{Theoretically expected $e$-like events (without
                     $\nu$ oscillations)}},\\ 
{\cal R}_\mu &\equiv& \frac{\mbox{Experimentally observed $\mu$-like events}}
                    {\mbox{Theoretically expected $\mu$-like events (without
                     $\nu$ oscillations)}},
\eeq
the Super-Kamiokande data reveals a significant $L/E$ dependence for
${\cal R}_\mu$, while ${\cal R}_e$ is consistent with unity with no $L/E$
dependence.
In this Letter, following an earlier study \cite{dva}, we establish that this
$L/E$ flatness of ${\cal R}_e$, consistent with ${\cal R}_e =1$,
constrains the neutrino mixing matrix dramatically.
All resulting neutrino-oscillation matrices are equivalent to each other,
under appropriate redefinitions of the underlying mass eigenstates.
They all carry the property that the expectation values of the muon and tau
neutrino masses be equal.
Moreover, the mixing matrices derived in this Letter naturally lead to a
consistency between the mixing angles observed by LSND and Super-Kamiokande.
This circumstance makes us suspect that the solar neutrino anomaly points
towards a richer phenomenology, perhaps beyond neutrino masses, in the
physics of neutrino oscillations.
%
%
\section{Constraints on the Neutrino-Oscillation Matrices}

Let us assume that at the top of the atmosphere, at $t=0$, the number of
$\nu_e$ and $\nu_\mu$ produced is $N_e$ and $N_\mu$, respectively.
Although both neutrinos and antineutrinos are produced in both flavours, we
shall use the terms ``electron neutrinos'' and ``muon neutrinos'' loosely,
to include both the $\nu$ and $\bar\nu$.
In general, the ratio of $\nu_\mu$ to $\nu_e$ neutrinos,
\beq
r = \frac{N_\mu}{N_e}
\eeq
is a function of energy.
However, for the relevant energy range in Super-Kamiokande, it may be assumed
constant (as shall be done in this Letter).
Within a detector at a distance $L\simeq t$ from the production point, the
number of electron neutrinos, $N_e'$, is given by
\beq
N_e' = N_e P_{ee} + N_\mu P_{\mu e},
\eeq
where $P_{ee}$ and $P_{\mu e}$ are the neutrino oscillation probabilities
$P(\nu_e \to \nu_e)$ and $P(\nu_\mu \to \nu_e)$, respectively.
Assuming that the underlying mass eigenstates are relativistic~\cite{dva_tg},
the neutrino oscillation probability $P(\nu_\ell \to \nu_{\ell '})$ takes the
form
\beq
&&P(\nu_\ell \to \nu_{\ell '}) = \delta_{\ell \ell '}\nonumber\\
&& -\, 4 \; Re(U_{\ell '1}U_{\ell 1}^*U_{\ell '2}^*U_{\ell 2})
\sin^2 (\varphi_{12})
   +   2 \; Im(U_{\ell '1}U_{\ell 1}^*U_{\ell '2}^*U_{\ell 2})
\sin (2 \varphi_{12})\nonumber\\
&& -\, 4 \; Re(U_{\ell '1}U_{\ell 1}^*U_{\ell '3}^*U_{\ell 3})
\sin^2 (\varphi_{13})
   +   2 \; Im(U_{\ell '1}U_{\ell 1}^*U_{\ell '3}^*U_{\ell 3})
\sin (2 \varphi_{13})\nonumber\\
&& -\, 4 \; Re(U_{\ell '2}U_{\ell 2}^*U_{\ell '3}^*U_{\ell 3})
\sin^2 (\varphi_{23})
   +   2 \; Im(U_{\ell '2}U_{\ell 2}^*U_{\ell '3}^*U_{\ell 3})
\sin (2 \varphi_{23}).  
\end{eqnarray}
The kinematic phases $\varphi_{ij}$, which appear in the above expression,
are defined as
\beq
\varphi_{ij} = 1.27 \Delta m_{ij}^2 \frac{L}{E}.
\eeq
In the above equation, $E$ is the neutrino energy (expressed in MeV), $L$ is
the distance between the generation point and the detection point (expressed
in meters), and $\Delta m_{ij}^2 \equiv m_i^2-m_j^2$ (expressed in eV$^2$).
For Dirac neutrinos, the matrix $U$ is a $3\times3$ unitary matrix,
parameterized in terms of three mixing angles $(\theta,\beta,\psi)$ and a
CP-violating phase, $\delta$:
\begin{equation}
U = \left(
\begin{array}{ccc}
c_\theta c_\beta             &
s_\theta c_\beta             &
         s_\beta
\\
- \,\, c_\theta s_\beta s_\psi e^{i\delta} - s_\theta c_\psi &
       c_\theta c_\psi - e^{i\delta} s_\theta s_\beta s_\psi &
       c_\beta s_\psi e^{i\delta}
\\
- \,\, c_\theta s_\beta c_\psi + s_\theta s_\psi e^{-i\delta} &
- \,\, s_\theta s_\beta c_\psi - c_\theta s_\psi e^{-i\delta} &
       c_\beta c_\psi
\end{array}
\right)
\end{equation}
in the Maiani representation~\cite{maiani}.
We use the abbreviated notations $c_\theta=\cos\theta$, $s_\theta=\sin\theta$,
etc., and we shall henceforth set the CP-violating phase $\delta=0$.
Assuming that the Super-Kamiokande data for electron-like events,
\beq
{\cal R}_e = \frac{N_e'}{N_e},
\eeq
is unity over its relevant $L/E$ range (an assumption which is certainly
valid within the systematic and statistical errors) implies that
\beq
P_{ee} + r P_{\mu e} = 1.
\eeq
Furthermore, from the unitarity condition, we have
\beq
P_{ee} + P_{e \mu} + P_{e \tau} = 1.
\eeq
Consequently, for a vanishing CP-violating phase
$\delta=0$ one has $P_{e \mu}=P_{\mu e}$, and we obtain
\beq
(r-1) P_{e \mu} = P_{e \tau}.
\eeq
Using the explicit expressions for the oscillation probabilities $P_{e \mu}$
and $P_{e \tau}$ yields
\beq
U_{e1} U_{e2} \left[(r-1)U_{\mu1}U_{\mu2} - 
U_{\tau1}U_{\tau2}\right] \sin^2 (\varphi_{12}) & + & \nonumber \\
U_{e1} U_{e3} \left[(r-1)U_{\mu1}U_{\mu3} - 
U_{\tau1}U_{\tau3}\right] \sin^2 (\varphi_{13}) & + & \nonumber \\
U_{e2} U_{e3} 
\left[(r-1)U_{\mu2}U_{\mu3} - 
U_{\tau2}U_{\tau3}\right] \sin^2 (\varphi_{23}) & = & 0.
\eeq
Since this condition should hold for all relevant values of $L/E$,
we obtain the following system of three equations with three unknowns 
($\theta$, $\beta$, and $\psi$):
\beq
U_{e1} U_{e2} \left[(r-1)U_{\mu1}U_{\mu2} - U_{\tau1}U_{\tau2}\right] & 
= & 0, \label{ue1ue2}\\
U_{e1} U_{e3} \left[(r-1)U_{\mu1}U_{\mu3} - U_{\tau1}U_{\tau3}\right] & 
= & 0, \label{ue1ue3}\\
U_{e2} U_{e3} \left[(r-1)U_{\mu2}U_{\mu3} - U_{\tau2}U_{\tau3}\right] & 
= & 0. \label{ue2ue3}
\eeq
In the following we shall first investigate the possible solutions of this
system of equations.
We then obtain the advertised result on the neutrino-mass degeneracy.
Finally, we briefly study the compatibility of the resulting
neutrino-oscillation mixing matrix for the LSND and Super-Kamiokande
experiments, as well as its consequences for the solar neutrino deficit.
%
%
\section{The Resulting Neutrino-Oscillation Mixing Matrices}

One possible class of solutions to the system of equations above is given by
requiring that the expressions in the square brackets in
Eqs.(\ref{ue1ue2})--(\ref{ue2ue3}) be zero simultaneously.
However, this leads to a rather uninteresting class of mixing matrices,
namely the unit matrix and others that are basically equivalent to it,
under appropriate redefinitions of the mass eigenstates.
Hence, the problem of the $L/E$ flatness in ${\cal R}_e$ is trivially solved,
but there would be no neutrino oscillations either.
This would contradict the existing data and we therefore discard such
solutions.

A more interesting class of solutions follows when one of $U_{e1}$,
$U_{e2}$, or $U_{e3}$ is zero.
This determines one of the mixing angles and two of the three equations are
trivially satisfied.
The remaining equation fully determines a second mixing angle.
We discuss this class of solutions below.
\subsection{The $U_{e1} = 0$ Case}

Since $U_{e1} = c_\theta c_\beta$, one solution to $U_{e1} = 0$ is
$c_\beta = 0$.
However, this implies that both $U_{e1}$ and $U_{e2}$ vanish identically,
which in turn means that there are no $\nu_e \to \nu_\mu$ and no
$\nu_e \to \nu_\tau$ oscillations.
We shall, therefore, discard this solution, as it is not of interest in the
context of the existing data.
The other solution to $U_{e1} = 0$ is $c_\theta = 0$, which implies
$s_\theta = \pm 1$.
Considering only the  $s_\theta > 0$ solution (as an illustration),
the mixing matrix becomes:
\begin{equation}
U = \left(
\begin{array}{ccc}
  0       &    c_\beta         & s_\beta        \\
- c_\psi  &  - s_\beta s_\psi  & c_\beta c_\psi \\
  s_\psi  &  - s_\beta c_\psi  & c_\beta c_\psi
\end{array}
\right).
\end{equation}
{}From Eq.(\ref{ue2ue3}) we have
\beq
s_\beta c_\beta [(r-1)s_\psi^2 - c_\psi^2] = 0,
\eeq
with two trivial solutions, $s_\beta = 0$ and $c_\beta = 0$.
They imply that $U_{e1} = 0$ and $U_{e3} = 0$, or $U_{e1} = 0$ and $U_{e2} = 0$,
respectively.
Both cases lead to no $\nu_e \to \nu_\mu$ and no $\nu_e \to \nu_\tau$
oscillations, and are discarded as discussed above.
More generally, however:
\beq
s_\psi & = & 1         /\sqrt{r}, \\
c_\psi & = & \sqrt{r-1}/\sqrt{r}.
\eeq
Notice that solutions are allowed for both $s_\psi < 0$ and $c_\psi < 0$,
but we shall restrict our 
(illustrative\footnote{The remaining $U$'s can be easily enumerated 
by the reader.})
discussion solely to the  $s_\psi > 0$ and $c_\psi > 0$ case.
At this point we explicitly set $r=2$ and thus the full mixing matrix becomes:
\beq
U = \left(
\begin{array}{ccc}
  0          &   c_\beta          & s_\beta          \\
- 1/\sqrt{2} & - s_\beta/\sqrt{2} & c_\beta/\sqrt{2} \\
  1/\sqrt{2} & - s_\beta/\sqrt{2} & c_\beta/\sqrt{2}
\end{array}
\right).
\eeq
\subsection{The $U_{e2} = 0$ Case}

Since $U_{e2} = s_\theta c_\beta$, one solution to $U_{e2} = 0$ is
$c_\beta = 0$, which is discarded as discussed above.
The $s_\theta = 0$ solution implies $c_\theta = \pm 1$, and considering only
the $c_\theta > 0$ solution, the mixing matrix becomes:
\begin{equation}
U = \left(
\begin{array}{ccc}
       c_\beta        &        0      & s_\beta        \\
- \,\, s_\beta s_\psi &        c_\psi & c_\beta s_\psi \\
- \,\, s_\beta c_\psi & - \,\, s_\psi & c_\beta c_\psi
\end{array}
\right).
\end{equation}
{}From Eq.(\ref{ue1ue3}) we have:
\beq
s_\beta c_\beta [(r-1)s_\psi^2 - c_\psi^2] = 0,
\eeq
and disregarding the trivial solutions $s_\beta = 0$ and $c_\beta = 0$, the
general mixing matrix yields (for $r = 2$):
\beq
U = \left(
\begin{array}{ccc}
  c_\beta          &   0          & s_\beta          \\
- s_\beta/\sqrt{2} &   1/\sqrt{2} & c_\beta/\sqrt{2} \\
- s_\beta/\sqrt{2} & - 1/\sqrt{2} & c_\beta/\sqrt{2}
\end{array}
\right).
\eeq
\subsection{The $U_{e3} = 0$ Case}

Since $U_{e3} = s_\beta$, $U_{e3} = 0$ implies simply that $s_\beta = 0$ and
thus $c_\beta = \pm 1$.
Considering only the  $c_\beta > 0$ solution, the mixing matrix becomes:
\beq
U = \left(
\begin{array}{ccc}
       c_\theta        &         s_\theta        & 0      \\
- \,\, s_\theta c_\psi &         c_\theta c_\psi & s_\psi \\
       s_\theta s_\psi &  - \,\, c_\theta s_\psi & c_\psi
\end{array}
\right).
\eeq

{}From Eq.(\ref{ue1ue2}) we have:
\beq
s_\theta c_\theta [s_\psi^2 - (r-1)c_\psi^2] = 0,
\eeq
and disregarding the trivial solutions $s_\theta = 0$ and $c_\theta = 0$,
the general solution leads to the following mixing matrix (for $r = 2$):
\beq
U = \left(
\begin{array}{ccc}
  c_\theta          &    s_\theta          & 0          \\
- s_\theta/\sqrt{2} &    c_\theta/\sqrt{2} & 1/\sqrt{2} \\
  s_\theta/\sqrt{2} &  - c_\theta/\sqrt{2} & 1/\sqrt{2}
\end{array}
\right).
\label{mixmat}
\eeq
%
%
\section{The Degenerate Muon and Tau Neutrino Masses}

Referring to the obtained neutrino-oscillation matrices $U$ above, and
taking note of the definition for the expectation value of the neutrino
masses (recall that for $\delta = 0$ all the $U_{\ell j}$ elements are real)
\beq
\langle m(\nu_\ell)\rangle \equiv \sum_j U^2_{\ell j}\,m_j,
\eeq
we immediately come to the general conclusion on the mass degeneracy of
the muon and tau neutrinos:
\beq
\langle m(\nu_\mu)\rangle = \langle m(\nu_\tau)\rangle.
\eeq
For the three cases enumerated above, one readily sees that the
``degenerate mass'' carries the values
\beq
&&\frac{1}{2}\left(m_1+s_\beta^2 m_2+c_\beta^2 m_3\right),\\
&&\frac{1}{2}\left(s_\beta^2 m_1+m_2+c_\beta^2 m_3\right),\\
&&\frac{1}{2}\left(s_\theta^2 m_1+c_\theta^2 m_2+m_3\right), 
\eeq
respectively.
The enumerated mixing matrices $U$ are immediately noted to be
equivalent to each other under redefinitions of the underlying mass eigenstates
$m_1$, $m_2$, and $m_3$.
%
%
\section{Implications for the Muon-Like Event Ratio: LSND versus
         Super-Kamiokande}

Apart from the mass degeneracy in the muon and tau neutrino masses,
a remarkable result that follows directly from the derived neutrino-oscillation
matrices is the consistency between the mixing angles observed by LSND and
Super-Kamiokande.
We briefly discuss this in the following.

The number of muon neutrinos at the Super-Kamiokande detector, $N_\mu'$, is
given by
\beq
N_\mu' = N_\mu P_{\mu \mu} + N_e P_{e\mu},
\eeq
and thus, the muon-like event ratio reads
\beq
{\cal R}_\mu = \frac{N_\mu'}{N_\mu} = P_{\mu \mu} + \frac{1}{r} P_{e\mu}.
\eeq
Without loss of generality, let us consider the mixing matrix given by
Eq.(\ref{mixmat}).
The $\nu_\mu$ survival probability, $P_{\mu\mu}$, and the $P_{e\mu}$
oscillation probability read
\beq
P_{\mu \mu} = 1 - s_\theta^2 c_\theta^2 \sin^2(\varphi_{12})
                -            s_\theta^2 \sin^2(\varphi_{13})
                -            c_\theta^2 \sin^2(\varphi_{23}),
\eeq
and
\beq
P_{e\mu} = 2 s_\theta^2 c_\theta^2 \sin^2(\varphi_{12}),
\label{Pemu}
\eeq
respectively.
Therefore, the Super-Kamiokande muon-like event ratio, ${\cal R}_\mu$, yields
\beq
{\cal R}_\mu = 1 - s_\theta^2 \sin^2(\varphi_{13})
                 - c_\theta^2 \sin^2(\varphi_{23}).
\label{Rmu}
\eeq
Here we have explicitly set $r = 2$.
At this point one cannot proceed any further without additional information on
either the mixing angle $\theta$, or the mass differences $\Delta m_{13}^2$ and
$\Delta m_{23}^2$.

This is the point where the LSND evidence comes into play.
As reported in Refs.~\cite{lsnddar,lsnddif}, the LSND experiment has obtained
evidence for neutrino oscillations in both the $\bar\nu_\mu \to \bar\nu_e$ and
$\nu_\mu \to \nu_e$ channels.
Although interpreted in terms of the simpler, two-generations neutrino mixing,
the allowed regions obtained by LSND help us gain further insight.
Within this framework, the $\nu_\mu \to \nu_e$ oscillation probability is
given by
\beq
P_{\mu e}^{LSND} = \sin^2 \left(2\Theta_{LSND}\right)
                   \sin^2 \left(\varphi_{12}\right),
\eeq
which is very similar to the expression in Eq.(\ref{Pemu}).
Indeed for the $\bar\nu_\mu \to \bar\nu_e$ and
$\nu_\mu \to \nu_e$ oscillation channels one may effectively identify
$ \sin^2 \left(2\Theta_{LSND}\right)$ with $2 s_\theta^2 c_\theta^2 =
\frac{1}{2} \sin^2 (2 \theta)$.
Therefore, a very small mixing angle $\Theta_{LSND}$, approximately of
${\cal O}(10^{-1})$ -- as indeed favored by the allowed regions indicated by
LSND -- implies a very small mixing angle $\theta$, also of
${\cal O}(10^{-1})$, in our formalism.
This in turn implies that the muon-like event ratio, ${\cal R}_\mu$, reads
\beq
{\cal R}_\mu = 1 - c_\theta^2 \sin^2(\varphi_{23}) + {\cal O}(10^{-2}),
\eeq
as opposed to simply
\beq
{\cal R}_\mu = 1 - \sin^2 \left(2\Theta_{SK}\right)
                   \sin^2 \left(\varphi_{23}\right),
\eeq
if ${\cal R}_\mu$ were to be expressed in the two-generations neutrino
oscillations formalism, where only $\nu_\mu \to \nu_\tau$ transitions are
allowed -- as interpreted by the Super-Kamiokande group.
Therefore, since $c_\theta \approx 1$, implies that
$\sin^2 (2\Theta_{SK}) \approx 1$ as well, as indeed reported in
Ref.~\cite{SuperK}.
%
%
\section{The Solar Neutrino Deficit}

If the Super-Kamiokande/LSND consistency is firmly established by future
experiments, then the physics of neutrino oscillations shall be found not only
to contain massive neutrinos, but may also point towards new physics.
This arises from the long-standing solar neutrino deficit, as measured by a
variety of experiments, with different sensitivities and detection
techniques~\cite{Homestake,Kamiokande,Sage,Gallex,SkSol}.
Within the framework of neutrino oscillations, the ratio of measured to
predicted solar neutrinos, $R_e$, is simply given by the $\nu_e$ survival
probability, $P_{ee}$.
Using the mixing matrix in Eq.(\ref{mixmat}), this reads
\beq
P_{ee} = 1 - 4 \, s^2_\theta c^2_\theta \, \sin^2 (\varphi_{12}),
\eeq
with an underlying mass scale $\Delta m_{12}^2 = {\cal O}(1) \,\, \mbox{eV}^2$,
as indicated by the LSND experiment.
Consequently, the kinematic term $\sin^2 (\varphi_{12})$ effectively averages
out to 1/2.
Furthermore, since the mixing angle $\theta$ is of ${\cal O}(10^{-1})$, as we
have argued in the previous Section, the predicted solar neutrino ratio is
practically $R_e = 1$, i.e., no solar neutrino deficit.
This is obviously in disagreement with the measured solar neutrino ratio of
$R_e \approx 0.5$, as reported by the above-mentioned experiments.
A popular solution to the solar neutrino deficit conjectures the existence of
a sterile neutrino(s).
This may very well be the way nature is.
However, before the sterile neutrino solution is invoked, one must make a
fundamental observation that the flavour and mass measurements do not
commute~\cite{dva}.
This incompatibility of the flavour and mass measurements can lead to a
violation of the principle of equivalence, which in turn modifies the standard
neutrino oscillation phenomenology in a fundamental manner~\cite{ep1}.
Such a violation of the principle of equivalence will take us into new physics.
At the same time, this might provide an elegant solution to the solar neutrino
anomaly~\cite{dva,ep1,ep2,ep3,ep4,ep5}.
%
%
\section{Conclusions}

We conclude that $L/E$ flatness of the electron-like event ratio in the
Super-Kamiokande data on atmospheric neutrinos implies a mass degeneracy for
the muon and tau neutrino.
The obtained results support recent considerations on maximal
mixing, bi-maximal mixing, and the degenerate neutrino
masses~\cite{s,je,dk,af}.
More precise data on the discussed $L/E$ flatness would be most helpful to
settle the insights gained in this Letter.
If the $L/E$ flatness of ${\cal R}_e$ is firmly established in the future by
the data, then one would be able to severely constrain the theoretical models
for neutrino masses and neutrino oscillations.
The general cases considered by us require that one of the $U_{e j}$ vanishes.
This result is in agreement with the conclusions reached by several authors,
see e.g. Ref.~\cite{gkln}.
Since all neutrino-oscillation matrices obtained by us are physically
equivalent, we have arrived at a unique $3\times 3$ neutrino-oscillation mixing
matrix that depends only on one angle.
This matrix clearly shows the consistency between the mixing angles observed
by the LSND and Super-Kamiokande experiments.

We explicitly note that the constraint implied by the $L/E$ flatness of the 
Super-Kamiokande e-like event ratio, as contained in this Letter, is a generalized 
version of that contained in Ref.~\cite{dva}. However, this is not the
main purpose of this Letter. Within the standard three-neutrino oscillation
framework, the mixing matrices that we now obtain exhaust 
all possibilities  consistent with the Super-Kamiokande implied constraint. 
In particular, we show that the Super-Kamiokande inferred 
$\nu_\mu \leftrightarrow \nu_\tau$ 
oscillation (with a complete decoupling from the $\nu_e$ oscillations) is too 
strong a conclusion. The class of solutions consistent with the Super-Kamiokande data, 
as systematically derived and analyzed here, is significantly richer and in particular 
leaves important room for oscillations away from, and into, the $\nu_e$ channel.
In Ref.~\cite{dva} only a very specific solution was obtained. 
This Letter obtains a class of new non-trivial and physically interesting solutions.
In particular, the LSND and Super-Kamiokande compatibility (as contained in bins 
where $L/E$ for LSND is close to that for Super-Kamiokande) emerges as a significant 
new result. With this compatibility established, it should now be clear to the
neutrino-oscillation community that Super-Kamiokande and LSND have an overlapping
and mutually consistent regime in the neutrino-oscillation parameter
space. Thus, either something in the non-overlapping regime of the Super-Kamiokande 
and LSND results must change to accommodate the solar neutrino anomaly -- or,
we must accept seriously that some new physics is hinted at.

%


\end{document}